# The first observation of effect of oscillation in Neutrino-4 experiment on search for sterile neutrino (continuation)

A.P. Serebrov[1], R.M. Samoilov[1], V.G. Ivochkin[1], A.K. Fomin[1], A.O. Polyushkin[1], V.G. Zinoviev[1],
P.V. Neustroev[1], V.L. Golovtsov[1], A.V. Chernyj[1], O.M. Zherebtsov[1], M.E. Chaikovskii[1], M.E. Zaytsev[1, 3],
A.A. Gerasimov[1], A.L. Petelin[2], A.L. Izhutov[2], A.A. Tuzov[2], S.A. Sazontov[2], M.O. Gromov[2],
V.V. Afanasiev[2],

1. *NRC "KI" Petersburg Nuclear Physics Institute, Gatchina, Russia*
2. *JSC "SSC Research Institute of Atomic Reactors", Dimitrovgrad, Russia*
3. *Dimitrovgrad Engineering and Technological Institute MEPhI, Dimitrovgrad, Russia*

E-mail: serebrov_ap@pnpi.nrcki.ru

## Abstract

We present the results of the Neutrino-4 experiment on search for a sterile neutrino. The experiment has been carried out on the SM-3 reactor having a compact active zone of 42x42x35cm$^3$ and operating on the highly enriched uranium-235 at 90 MW thermal power. We report the results of the Neutrino-4 experiment of measurements of reactor antineutrino flux and spectrum dependence on the distance in the range 6-12 meters from the center of the reactor core. Using the measured spectrum and the distance dependence of antineutrino flux, we performed the model independent analysis of restrictions on the oscillation parameters $\Delta m_{14}^2$ and $\sin^2 2\theta_{14}$. The method of coherent addition of results of measurements is proposed. It allows us to directly observe the effect of oscillations. We observed the oscillation effect at CL 3.5$\sigma$ in the vicinity of $\Delta m_{14}^2 \approx 7.26 \text{eV}^2$ and $\sin^2 2\theta_{14} \approx 0.38$. Combining the result of the Neutrino-4 experiment and the result of the gallium anomaly effect we obtained value $\sin^2 2\theta_{14} \approx 0.35\pm0.07$ (5.0$\sigma$). The analysis of systematics effects is presented. Comparison with results of other experiments is presented. Future prospect of the experiment is discussed. It is necessary to notice that obtained values $\sin^2 2\theta_{14} \approx 0.35\pm0.07$ (5.0$\sigma$) and $\Delta m_{14}^2 = (7.3 \pm 0.7)\text{eV}^2$ allow make assessment on the mass of a neutrino: $m_\beta \approx 0.8$ eV.

## 1. Introduction

At present, there is a widely spread discussion on the possible existence of a sterile neutrino. It is assumed, that due to possible reactor antineutrino transition to the sterile state, the oscillation effect at short reactor distances can be observed [1,2]. Moreover, a sterile neutrino can be considered as a candidate for the dark matter.

Ratio of observed/predicted antineutrino flux in various reactor experiments is estimated as 0.934 ± 0.024 [3]. The effect is 3 standard deviations. This, however, is not yet sufficient to have confidence in existence of the reactor antineutrino anomaly. The method of comparison of measured antineutrino flux from the reactor with expected calculated value requires precise estimation of both antineutrino flux from reactor and neutrino detector efficiency. This is the method of absolute measurements.

The hypothesis of oscillation can be verified by direct measurement of the antineutrino flux and spectrum vs. distance at short 6 – 12m distances from the reactor core. This is the method of relative measurements and it can be more precise. A detector is supposed to be movable and spectrum sensitive. Our experiment focuses on the task of exploring the possible existence of a sterile neutrino at a certain confidence level or refuting this hypothesis. To detect oscillations to a sterile state, it would be indicative to observe the deviation of flux-distance relation from 1/L$^2$ dependence. If such a process does occur, it can be described at short distances by the equation:

$$P(\bar{\nu}_e \to \bar{\nu}_e) = 1 - \sin^2 2\theta_{14} \sin^2 \left(1.27 \frac{\Delta m_{14}^2 [\text{eV}^2] L[m]}{E_{\bar{\nu}}[\text{MeV}]}\right) \quad (1)$$

where $E_{\bar{\nu}}$ is antineutrino energy, with oscillation parameters $\Delta m_{14}^2$ and $\sin^2 2\theta_{14}$ being unknown. For the experiment to be conducted, one needs to carry out measurements of the antineutrino flux and spectrum as near as possible to a practically point-like antineutrino source.

We have studied several options for carrying out our new experiments at research reactors in Russia. The research reactors should be employed for performing such experiments, since they possess a compact reactor core, so that a neutrino detector can be placed at a small distance from it. Unfortunately, a research reactor beam hall usually has quite a large background of neutrons and gamma quanta, which makes it difficult to carry out low background experiments. Due to some peculiar characteristics of its construction, reactor SM-3 provides the most favorable conditions to search for neutrino oscillations at short distances [4, 5]. However, the SM-3 reactor, as well as other research reactors, is located on the Earth surface, hence, the cosmic background is the major difficulty in considered experiment.

## 2. Detector design

Detector scheme with active and passive shielding is shown in Fig. 1. The liquid scintillator detector has volume of 1.8 m$^3$ (5x10 sections 0.225x0.225x0.85м$^3$, filled to the height of 70 cm). Scintillator with gadolinium concentration 0.1% was used to detect inverse beta decay (IBD) events $\bar{\nu}_e + p \rightarrow e^+ + n$. The method of antineutrino registration is to select correlated pare of signals: prompt positron signal and delayed signal of neutron captured by gadolinium.

The neutrino detector active shielding consists of external and internal parts relative to passive shielding. The internal active shielding is located on the top of the detector and under it. The detector has a sectional structure. It consists of 50 sections – ten rows with 5 sections in each. The first and last detector rows were also used as an active shielding and at the same time as a passive shielding from the fast neutrons. Thus, the fiducial volume of the scintillator is 1.42 m$^3$. For carrying out measurements, the detector has been moved to various positions at the distances divisible by section size. As a result, different sections can be placed at the same coordinates with respect to the reactor except for the edges at closest and farthest positions.

Construction of a multi section system was aimed at using additional criteria for selection of neutrino events. The main problem of the experiment on the Earth's surface is fast neutrons from cosmic radiation. The elastic scattering of fast neutrons easily imitates an IBD, which is an indicative reaction of antineutrino. Registration of the first (start or prompt) signals from recoil protons imitates registration of a positron. The second (stop or delayed) signal arises in both cases when a neutron is captured by gadolinium. The difference between these prompt signals is in appearance of two gamma quanta, propagating in opposite directions with energy 511 keV each, produced in annihilation of a positron from IBD process. The recoil proton track with high probability is located within the size of one detector section, because its track length is about ~1 mm. Positron free path in an organic scintillator is ~5 cm, hence if its signal is detected in a section then 511 keV gamma-quanta could be detected in an adjacent section.

Monte Carlo calculations has shown that 63% of prompt signals from neutrino events are recorded within one section and only 37% of events has signal in another section [6]. In our measurements, the signal difference at the reactor ON and OFF has ratios of double and single prompt events integrated over all distances $(37\pm4)\%$ and $(63\pm7)\%$ respectively. This ratio allows us to interpret the recorded events as neutrino events within current experimental accuracy. Unfortunately, a more detailed analysis of that ratio cannot be performed due to low statistical accuracy. Yet, it should be noted, that the measurements of fast neutrons and gamma fluxes in dependence on distance and reactor power were made before installing the detector into passive shielding [6, 7]. Absence of noticeable dependence of the background on both distance and reactor power was observed. As a result, we consider that difference in reactor ON-OFF signals appears mostly due to antineutrino flux from operating reactor. That hypothesis is confirmed by the given above ratio of single and multi-section prompt signals typical especially for neutrino events. The fraction of fast neutrons at reactor activation does not exceed 3% of the neutrino effect.

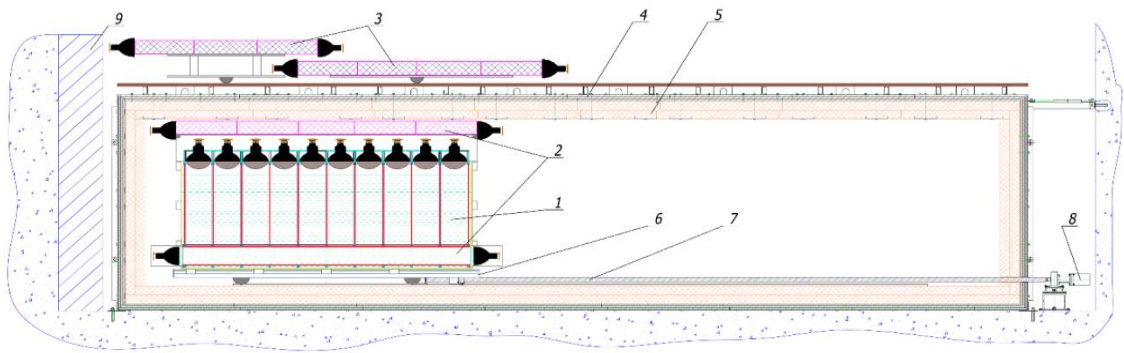

Fig. 1. General scheme of an experimental setup. 1 – detector of reactor antineutrino, 2 – internal active shielding, 3 – external active shielding (umbrella), 4 – steel and lead passive shielding, 5 – borated polyethylene passive shielding, 6 – moveable platform, 7 – feed screw, 8 – step motor, 9 – shielding against fast neutrons from iron shot.

## 3. Measurement results - scheme of reactor operation and detector movements

The first measurements with the detector have started in June 2016 and was continued till June 2018. Measurements with the reactor ON were carried out for 480 days, and with the reactor OFF- for 278 days. In total, the reactor was switched on and off 58 times.

The scheme of reactor operation and detector movements is shown in Fig. 2 at the top. The measurements of the background (OFF) and measurements with reactor in operation mode (ON) are carried out within the exposure period at single detector position. A reactor cycle is 8-10 days long. Reactor shutdowns are 2-5 days long and usually alternates (2-5-2-...). The reactor shutdowns at summer for a long period for scheduled preventive maintenance. The movement of the detector to the next measuring position occurs in the middle of reactor operational cycle. The stability of the results of measurements is characterized by distributions of ON-OFF difference fluctuations normalized on their statistical uncertainties, in measurements within one period. The distribution is shown in Fig. 2 at the bottom.

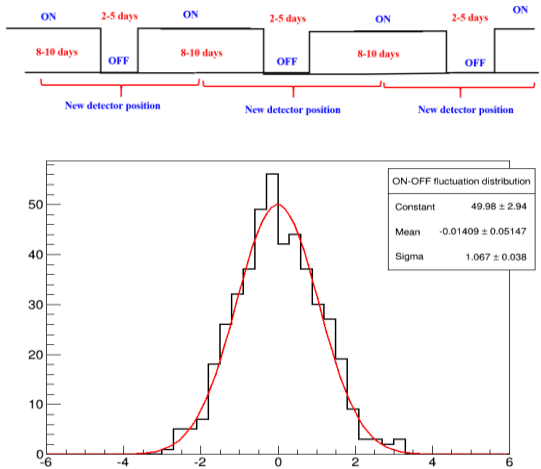

Fig. 2. Top - scheme of detector operation; bottom - the distribution of deviations from average value of correlated events rates differences (ON-OFF) normalized on its statistical uncertainties.

That distribution has the form of normal distribution, but its width exceeds unit by 7%. This is a result of additional dispersion which appears due to fluctuations of cosmic background and impossibility of simultaneous measurements of the effect and background. Since the measurements of the background carried out during the annual scheduled reactor repair works, when the reactor is stopped for a month, are added to total obtained data, then total additional dispersion, which is a result of background measurements, increases up to 9%.

## 4. Measurements 2016-2018. Dependence of antineutrino flux on the distance to the reactor core

Results of measurements of the difference in counting rate of neutrino-like events for the detector are shown in Fig. 3, as dependence of antineutrino flux on the distance to the reactor core. Fit of an experimental dependence with the law $A/L^2$ yields satisfactory result. Goodness of that fit is 81%.

Corrections for finite size of reactor core and detector sections are negligible – 0.3%, and correction for difference between detector movement axes and direction to center of reactor core is also negligible – about 0.6%.

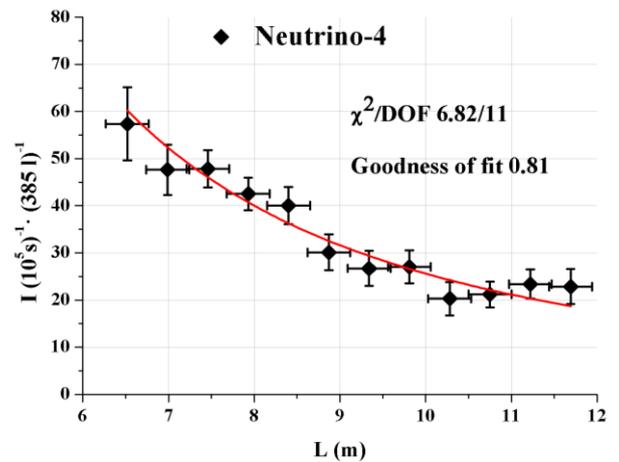

Fig.3. Experimental results fitted with function $A/L^2$.

## 5. Energy calibration of the detector

Energy calibration of the detector was performed with γ-quanta source and neutron source (Fig.4) ($^{22}$Na by lines 511 keV and 1274 keV, by line 2.2 MeV from reaction np-dγ, by gamma line 4.44 MeV from Pb-Be source, and also by total energy of gamma quanta 8 MeV from neutron capture in Gd. Calibration for energy 8 MeV is difficult because probability of registration of three gamma quanta in one section is small) [7].

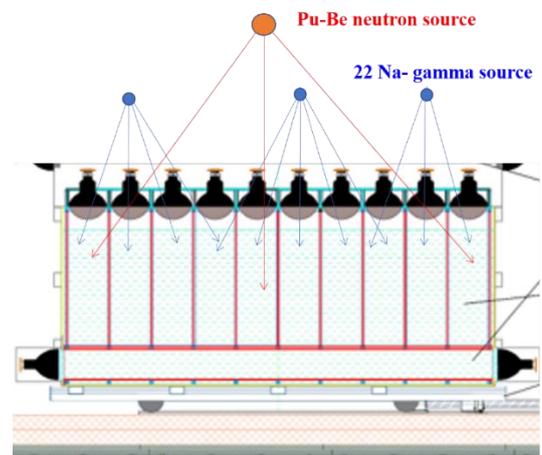

Fig 4. Energy calibration of the detector

These calibration spectra are shown in Fig. 5 (detailed) and in Fig. 6. Fig. 7 demonstrates linearity of calibration dependence. As a result, spectrum of prompt signals registered by detector was measured. Its connection with antineutrino energy is determined by equation: $E_{promt} = E_{\bar{\nu}} - 1.8 \text{ MeV} + 2 \cdot 0.511 \text{ MeV}$, where $E_{\bar{\nu}}$ - antineutrino energy, 1.8 MeV – energy threshold of IBD, and $2 \cdot 0.511$ MeV corresponds to annihilation energy of a positron.

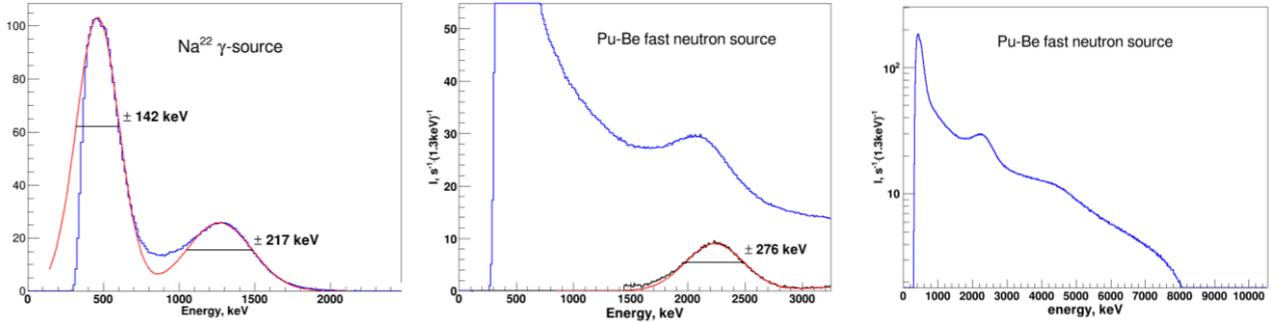

Fig. 5. Calibration in details.

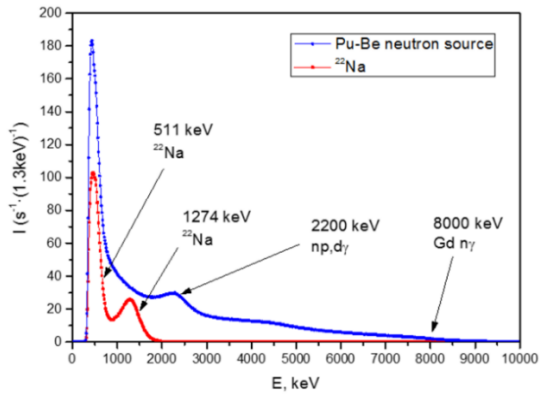

Fig. 6 The results of detector calibration.

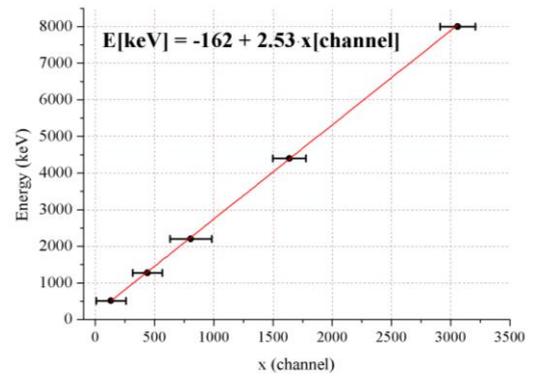

Fig. 7. Linearity of calibration dependence.

## 6. Experimental spectrum of prompt signals

However, detector efficiency has to be taken into account and obtained spectrum should be compared with the one simulated using MC model of the detector. An example of such comparison is shown in Fig.8, where we present experimental spectrum of prompt signals averaged over all distances for better statistical accuracy and MC spectrum of prompt signals, obtained using spectrum of $^{235}$U [1] and with considering thresholds of experimental signals.

A discrepancy of experimental and calculated spectra is observed at 3MeV. Spectra are normalized to the experimental one. Their ratio is shown at Fig.9.

The ratios of the experimental spectra of prompt signals averaged over three distance ranges (~2m) with centers in points 7.3 m, 9.3 m and 11.1 m. to the spectrum simulated with MC calculations are shown in Fig. 9a. Averaged over all distances ratio and its polynomial fit (red curve) are shown in Fig. 9b. It should be noted that deviation of experimental spectrum from calculated one is equal, within experimental accuracy, for different distances. Red curve fits all distance points equally well. Goodness of fits are 77%, 78% and 68% for three distances 7.3m, 9.3m and 11.1m correspondingly.

So-called "bump" in 5 MeV area is also observed just as in other experiments [8-12], but its amplitude is larger than in experiments at nuclear power plants. If it is connected with $^{235}$U, as assumed in works [13-15], then it could be explained by a high content of $^{235}$U (95%) at SM-3 reactor in distinction from effective fission fraction of $^{235}$U 56% [11] or 65% [8,9,12] at different industrial reactors.

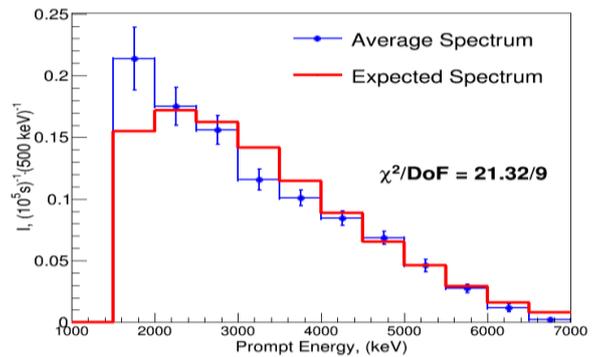

Fig. 8 The spectrum of prompt signals in the detector for a total cycle of measurements summed over all distances (average distance — 8.6 meters). The red line shows the Monte-Carlo simulation with neutrino spectrum for $^{235}$U [1], as the SM-3 reactor works on highly enriched uranium.

Thus, calculations of reactor flux can be one of the possible reasons for the discrepancy. Taking into consideration 0.934 deficiency for an experimental antineutrino flux with respect to the calculated one, we should discuss not the «bump» in 5 MeV area, but the «hole» in 3 MeV area. However, one should take into account influence of oscillations with high $m_{14}^2$ because we use 2m interval in analysis.

Using such averaging, if $\Delta m_{14}^2 > 5 eV^2$ then spectrum would be suppressed by factor $1 - 0.5 \sin^2 2\theta_{14}$ starting from low energies. Lastly, we should also consider possibility of systematic errors in calibration of energy scale or Monte-Carlo calculations of prompt signal spectrum in low energy region. There is a problem of precise registration of annihilation gamma energy (511 keV) in adjacent sections. Thus, energy point 1.5 MeV is the most problematic one.

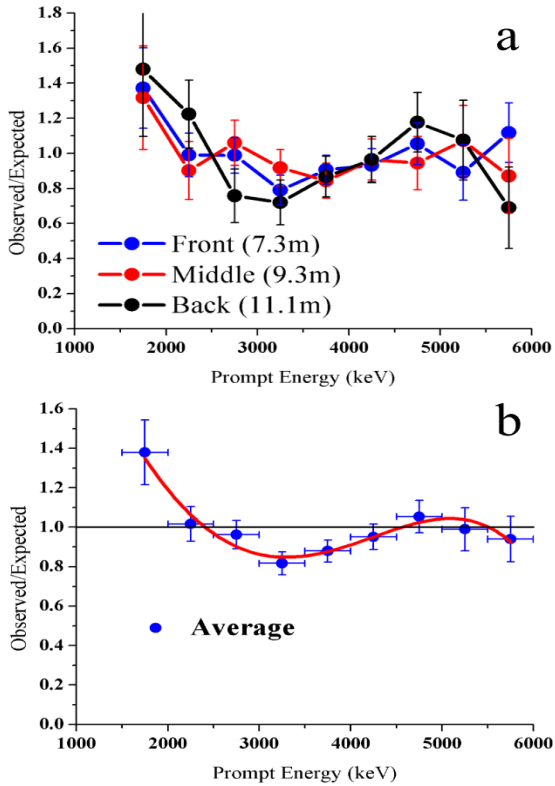

Fig. 9. a – The ratio of an experimental spectrum of prompt signals to the spectrum, expected from MC calculations for 3 ranges (~2m) with centers 7.3m, 9.3m and 11.1m b – polynomial fit of results averaged by distance (red curve)

## 7. Experimental results - the matrix of measurements of the antineutrino flux dependence on distance and energy

The results of experimental measurements of the antineurino flux dependence on a distance and energy of antineurino can be presented in the form of a matrix, which contains 216 elements, where N(i, k) is difference of ON - OFF account for i-th interval of energy and for k-th distance from reactor core. The energy spectrum is divided into 9 intervals of 500 keV according to the energy resolution of the detector elements. The detail of antineurino flux measurement corresponds to the cell step of 23cm. In total there are 24 positions of antineurino flux measurement from 6.4m to 11.9m.

## 8. Analysis of the experimental result

As previously shown, there is a problem with discrepancy between the experimental and calculated spectra.

Therefore, method of the analysis of the experimental data should not rely on precise knowledge of the spectrum. One can carry out model independent analysis using equation (2), where the numerator is the rate of antineutrino events per $10^5$s with a correction to geometric factor $L^2$ and denominator is the antineutrino events rate averaged over all distances:

$$(N_{i,k} \pm \Delta N_{i,k})L_k^2 / [K^{-1} \sum_{k}^{K} (N_{i,k} \pm \Delta N_{i,k})L_k^2] =$$
$$= [1 - \sin^2 2\theta_{14} \sin^2(1.27 \Delta m_{14}^2 L_k / E_i)] / \quad (2)$$
$$[K^{-1} \sum_{k}^{K} [1 - \sin^2 2\theta_{14} \sin^2(1.27 \Delta m_{14}^2 L_k / E_i)]]$$

Equation (2) is model independent because the left part includes only experimental data $k = 1, 2, ... K$ for all distances in the range 6.4-11.9 m, $K = 24$; $i = 1, 2, ... 9$ corresponding to 500keV energy intervals in range 1.5MeV to 6.0MeV. The right part is the same ratio obtained within oscillation hypothesis. The left part is normalized to spectrum averaged over all distances; hence the oscillation effect is considerably averaged out in denominator if oscillations are frequent enough in considered distances range.

Using all 24 positions instead of 3 as we did before [7], we increase analysis sensitivity to high values of $\Delta m_{14}^2$. Averaging the results over 3 positions (2 meters each) one cannot observe oscillations with period less than 2 meters.

The results of the analysis of experimental data using equation (2) with $\Delta \chi^2$ method are shown in Fig 10.

The area of oscillation parameters colored in pink are excluded with CL more than 99.73% (>3σ). However, in area $\Delta m_{14}^2 = (7.34 \pm 0.1) eV^2$ and $\sin^2 2\theta_{14} = 0.44 \pm 0.14$ the oscillation effect is observed at CL 99% (3σ), and it is followed by a few satellites. Minimal value $\chi^2$ occurs at $\Delta m_{14}^2 \approx 7.34 eV^2$.

The satellites appear due to effect of harmonic analysis where in presence of noises along with base frequency we also can obtain frequencies equal to base frequency multiplied by integers and half-integers.

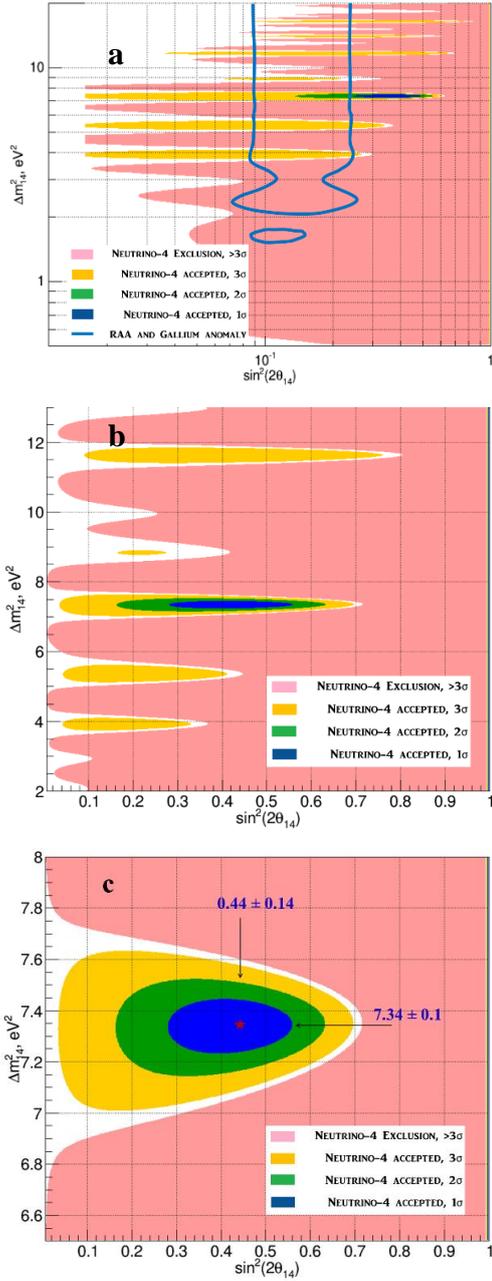

Fig. 10. a – Restrictions on parameters of oscillation into sterile state with 99.73% CL (pink), area of acceptable with 99.73% CL values of the parameters (yellow), area of acceptable with 95.45% CL values of the parameters (green), area of acceptable with 68.30% CL values of the parameters (blue). b – Area around central values in linear scale and significantly magnified, c – even further magnified central part.

The stability of the results of the analysis can be tested. Using the obtained experimental data $(N_{i,k} \pm \Delta N_{i,k})$ one can perform a data simulation using randomization with a normal distribution around $N_{i,k}$ with dispersion $\Delta N_{i,k}$. Applying this method, 60 virtual experiments were simulated with results lying within current experimental accuracy. One can carry out the analysis described above for virtual experiments and average results over all distributions. It was observed that exclusion area (pink area in Fig. 10a) coincide with experimental one and oscillation effect area is gathered around value $\Delta m_{14}^2 \approx 7.3 \text{eV}^2$.

Finally, one can simulate the experimental results with same accuracy but in assumption of zero antineutrino oscillations. Obtained result reveals that amplitude of perturbations in horizontal axes, i.e. values of $\sin^2 2\theta_{14}$, is significantly reduced. It signifies that big perturbations in Fig. 10a indicate an existence of the oscillation effect. Simulated experimental data distributions with same accuracy, but in assumption of zero oscillation allows us to estimate sensitivity of the experiment at CL 95% and 99%. Obtained estimations can be used to compare our results with other experiments.

## 9. Method of coherent summation of experimental data by L/E parameter.

Fig. 11 illustrates the method of coherent summation of experimental data by L/E parameter. We have to sum up experimental data at the same L/E parameter.

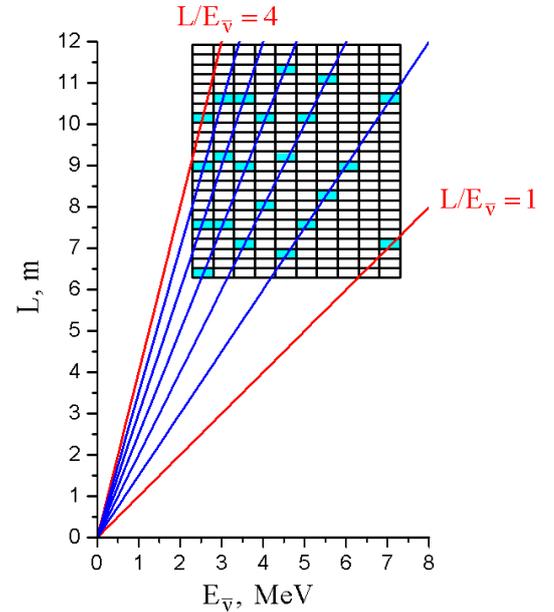

Fig. 11. Method of coherent summation of data by L/E parameter. Summing up experimental data at the same L/E parameter.

Since, according to equation (1), oscillation effect depends on ratio L/E, it is beneficial to make experimental data selection using that parameter. That method we call the coherent summation of the experimental results with data selection using variable L/E and it provides direct observation of antineutrino oscillation.

For this purpose, we used 24 distance points (with 23 cm interval) and 9 energy points (with 0.5MeV interval). The corresponding matrix is schematically shown in Fig. 11. The selection for left part of equation (2) (of total 216 points each 8 points are averaged) is shown in Fig. 12 with blue triangles.

Same selection for right part of equation (2) with most probable parameters $\Delta m_{14}^2 \approx 7.34 eV^2$ and $\sin^2 2\theta_{14} \approx 0.44$ is also shown in Fig.12 with red dots. The fit with such parameters has the goodness of fit 90%, while fit with a constant equal to one (assumption of no oscillations) has goodness of fit only 27%. It is important to notice that attenuation of sinusoidal process for red curve in area L/E > 2.5 can be explained by taken energy interval 0.5MeV. Considering the smaller interval 0.25MeV we did not obtain increasing of oscillation area of blue experimental, because of insufficient energy resolution of the detector in low energy region. Thus, the data obtained in region L/E > 2.5 do not influence registration of oscillation process. Using first 19 points in analysis, we obtained new $\chi^2$ and goodness of fit which are shown under the curve in Fig.13. In Fig. 12 and Fig.13 the vertical errors are statistical one, the horizontal errors correspond to the interval of averaging of data.

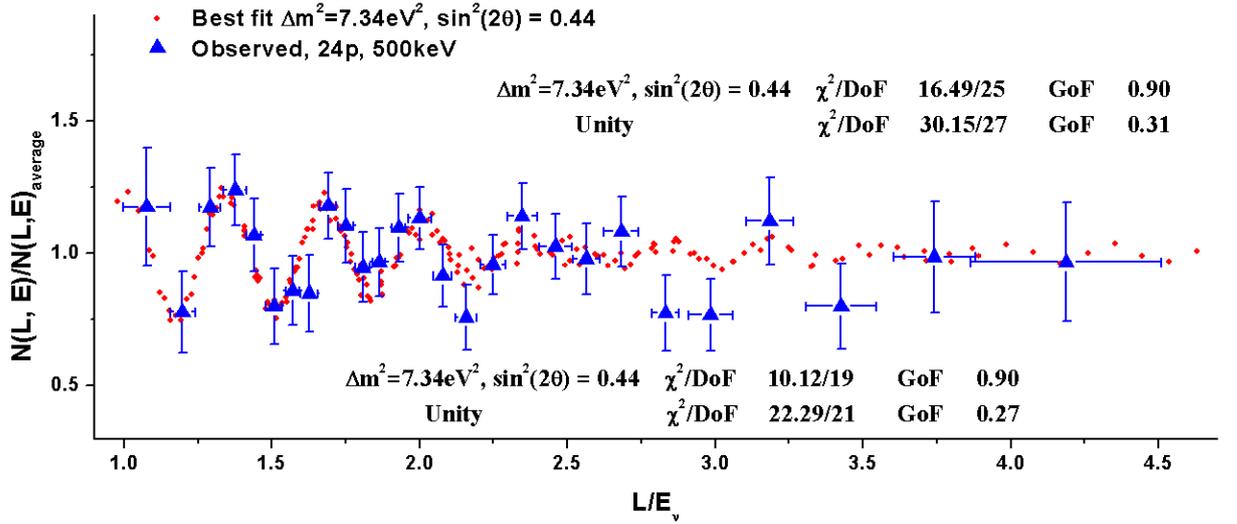

Fig.12. The coherent addition of the experimental result with data selection by variable L/E for direct observation of antineutrino oscillation. Comparison of left (blue triangles) and right (red dots, with optimal oscillation parameters) parts of equation (2).

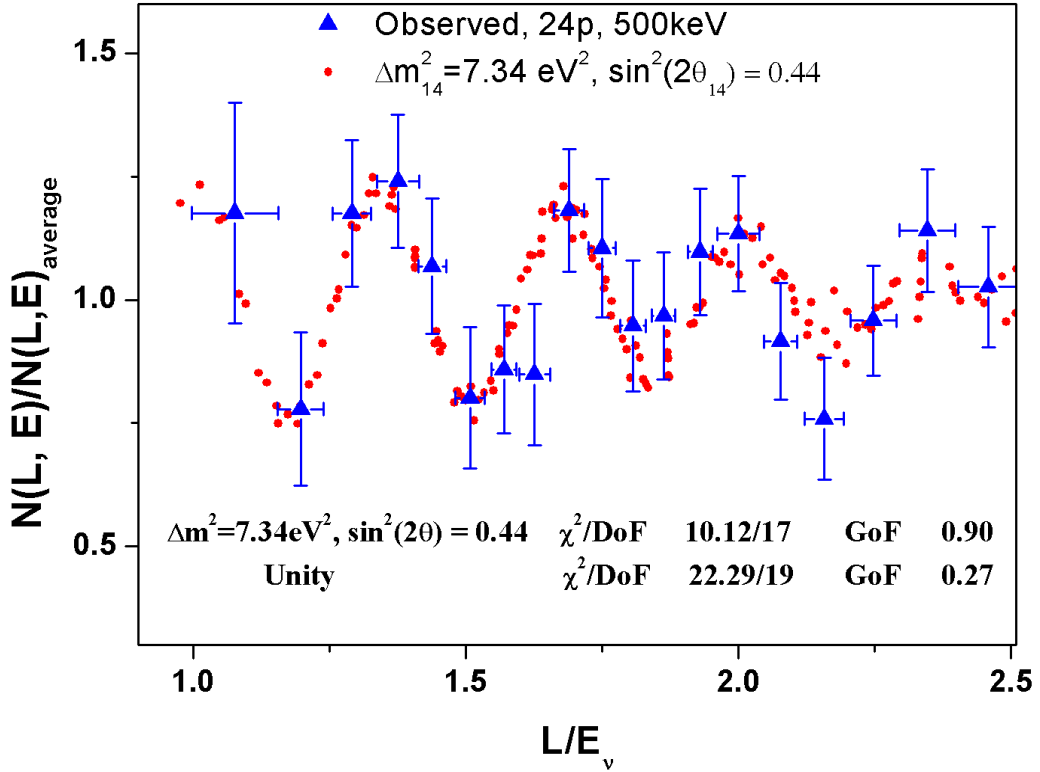

Fig.13. The most important part of effect of antineutrino oscillation in sterile neutrino in experiment Neutrino-4.

It should be noticed, that the product of expected spectrum (spectrum of $^{235}$U in assumption of no oscillations) and oscillation factor for each distance are integrated over intervals corresponding to energy intervals in left hand side (1.5MeV – 2MeV, 2 – 2.5MeV …).

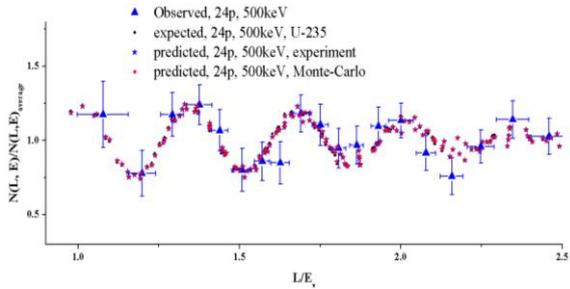

Fig. 14. Comparison of experimental data with expected forms of the dependences in assumption of various initial neutrino spectra. Black dots - the spectrum of $^{235}$U, blue stars - experimental spectrum averaged over all distances, red rhombuses - the results of Monte-Carlo simulation of neutrino spectrum for full-scale detector

However, as shown in Fig. 14, the resulting function of L/E is independent on the initial expected

## 10. Analysis of systematic effects

To carry out an analysis of possible systematic effects one should turn off antineutrino flux (reactor) and perform the same analysis of obtained data, which consist of signals of fast neutrons from cosmic rays. It is very important to mention that the spectrum of recoil protons is similar to the spectrum of positrons in inverse beta decay (IBD) events $\bar{\nu}_e + p \rightarrow e^+ + n$. The result of that analysis is shown in Fig.16 and it indicates the absence of oscillations in the analyzed area.

Correlated background (fast neutrons from cosmic rays) slightly decreases at farther distances from reactor due to inequality of concrete elements of the building, which comes out as linear decrease (red line) in Fig. 16a. It results in green zone at oscillation parameters, $\Delta m_{14}^2$, $\sin^2 2\theta_{14}$ plane, which has absolutely no connection with oscillation effect. The deviation of results from linear law, showed in Fig.16c, cannot be the reason of observation of oscillations effect. Thus, no instrumental systematic errors were observed.

The distances of detector movements are multiples of section size (23cm). All movements are controlled with laser distance measurer. The measurements were carried out at 10 detector positions in the way that the same distance from the reactor is measured with various detector rows. Spectra measured with various rows at same distance are averaged afterwards.

spectrum, hence with high accuracy one can consider that the energy spectrum is cancelled out in right hand side in (2). Also, number of energy bins and averaging step are chosen in convenient way. However, selection of the arbitrary values of the parameters would not results in any significant difference, as shown in Fig.15.

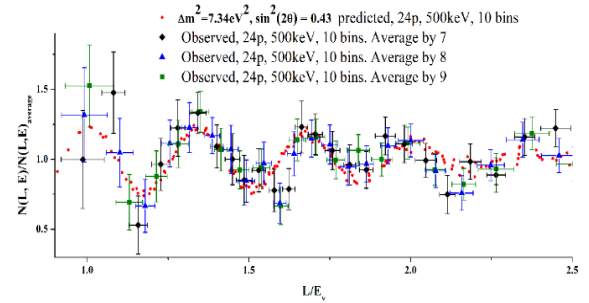

Fig. 15. The results of coherent summation with various averaging steps of energy spectrum in range 1.5 - 6.5 MeV.

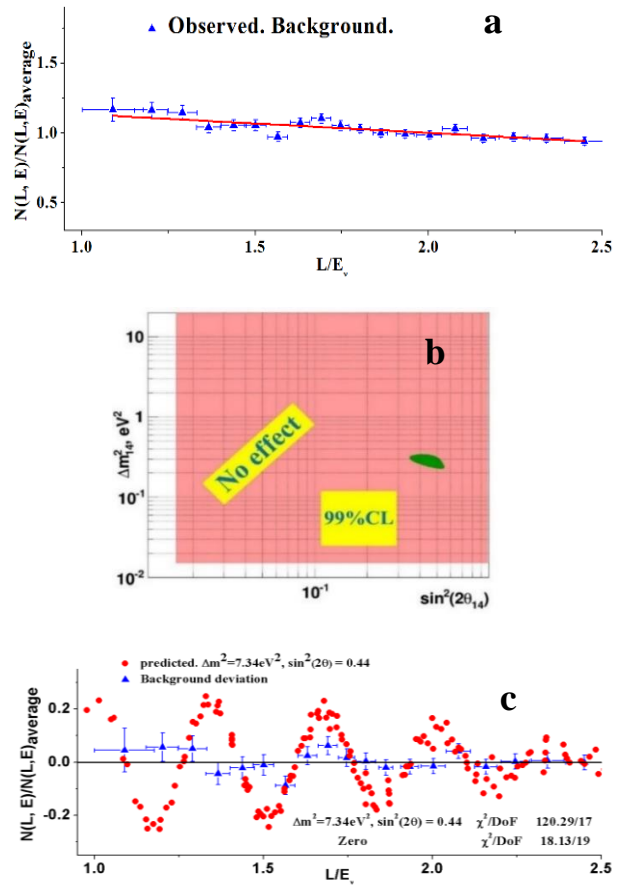

Fig.16. Analysis of data obtained with turned off reactor carried out to test on possible systematic effects: a - data analysis using coherent summation method, b - analysis of the results on oscillation parameters plane. c- dots corresponds to deviation of expected effect from the unit, triangles - deviation of background from the linearly decreasing trend at Fig. 16a.

Average distribution of prompt signal counts obtained in background measurements during the whole period of reactor stop is shown in Fig. 17 (top). It was mentioned before that cosmic background of fast neutrons in lab room is inhomogeneous due to the building structure. It appears as a slope of background dependence on L/E in Fig. 16a, and as the profile of that distribution (red line in Fig. 17 top). Therefore, to estimate how the detector inhomogeneity can affect the results, one should consider the deviation of counts from that profile, as shown in Fig. 17 (bottom). We should remind that first and last rows are not used for obtaining the final dependence on L/E and mean value of the deviation is ~ 8%.

To consider how differences in rows efficiencies affect the final results, one must take into account the averaging of spectra obtained with various rows at the same distance. Hence the relative contribution of each row must be calculated. In that approach the squared deviation from the mean value is ~ 2.5%, as shown in Fig. 18. It indicates that the influence of detector inhomogeneity on the L/E dependence is insignificant and cannot be the origin of oscillation effect.

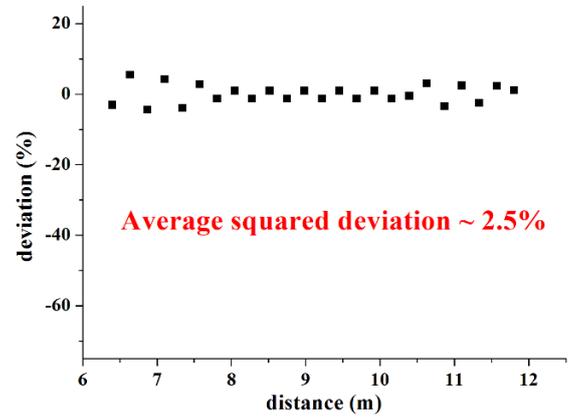

Fig. 18. Deviation of counts of correlated background of each distance from the reactor after averaging over rows from the mean value.

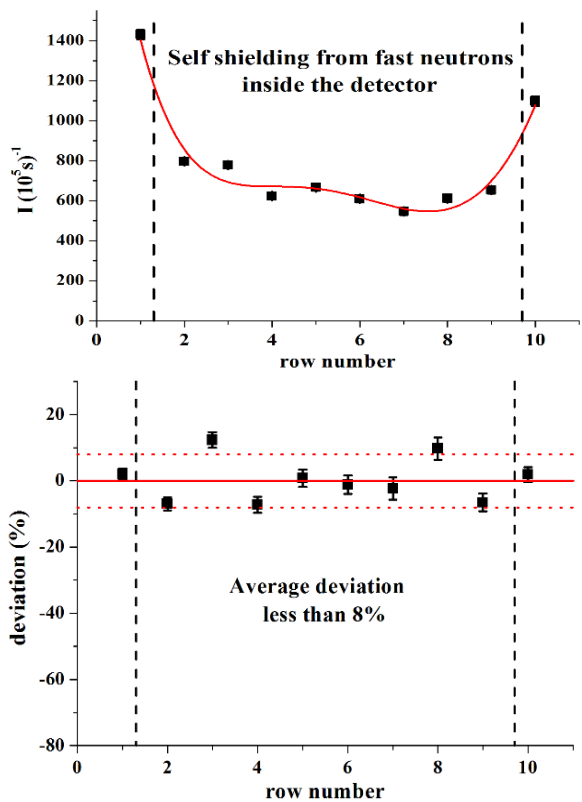

Fig.17. Average distribution of correlated background prompt signals in detector over all positions (top). Deviation average distribution of prompt signals from profile. Profile was caused by inhomogeneity of fast neutrons background in the lab room (bottom).

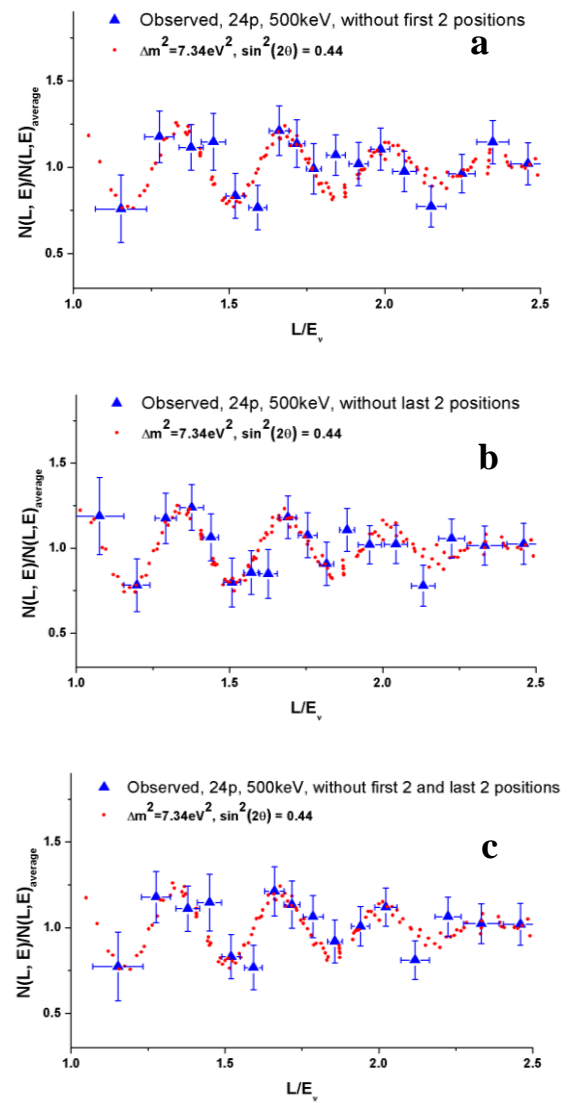

Fig. 19. a – the result of coherent summation in data analysis without first two distances; b – without last two distances; c – without first two and last two distances.

To provide an additional test one can exclude from the analysis the measurements made by second and third rows at the position closest to the reactor and by eighth and ninth rows at the farthest from the reactor position, for those are extreme positions and corresponding measurements are not averaged with any other rows.

The result of the test is shown in Fig. 19 where one can see that the oscillation effect remains, but the statistical accuracy decreases after data exclusion and CL reduced to ~2σ

## 11. Monte Carlo calculation

The red dots in Fig. 20 were obtained by Monte Carlo calculation. A source of antineutrino with geometrical dimensions of the reactor core 42x42x35cm$^3$ was played, as well as a detector of antineutrino taking into account its geometrical dimensions (50 sections of 22.5x22.5x85cm). The antineutrino spectrum of U$^{235}$ (though it did not matter since the energy spectrum of an antineutrino in the equation (2) is reduced) increased by function of oscillations $1 - \sin^2 2\theta_{14} \sin^2(1.27\Delta m_{14}^2 L_k/E_i)$ was used. The most important parameter in this simulation was the energy resolution of the detector, which was 500 keV. Fig. 20 (left) shows the relationship of the oscillation pattern to the energy resolution of the detector. The energy resolution oscillation curve was verified. The energy resolution of the 500 keV is best suited to the description of experimental data.

Fig. 20 (right) shows the model matrix of ratio $(N_{ik} \pm \Delta N_{ik})L_k^2/K\sum(N_{ik} \pm \Delta N)L_k^2$ for calculations, where $\Delta N_{ik}/N_{ik}$ is 1%. A picture of the process of oscillations on the plane (E, L) can be seen.

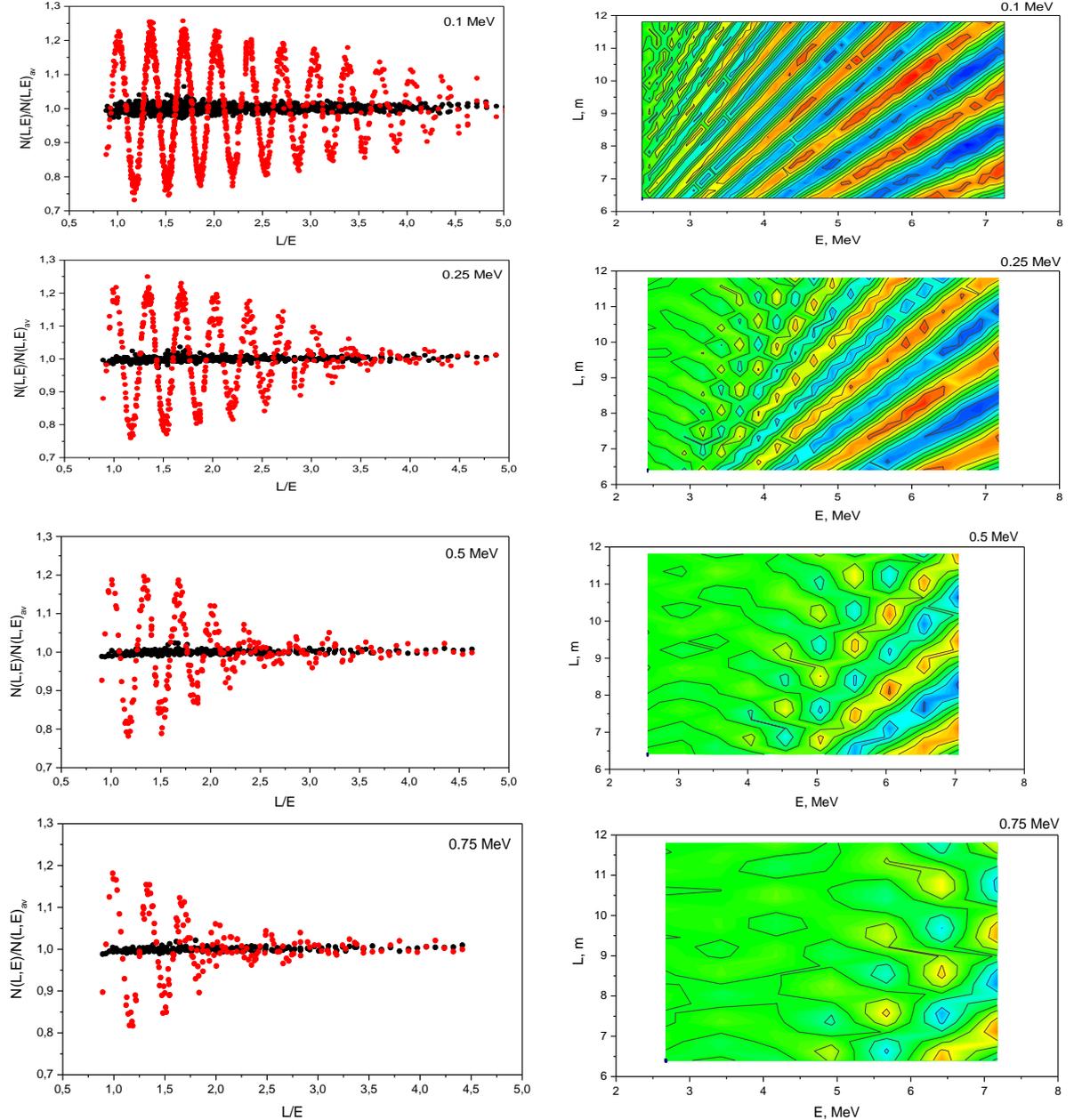

Fig.20. The model matrix of ratio $(N_{ik} \pm \Delta N_{ik})L_k^2/K\sum(N_{ik} \pm \Delta N)L_k^2$ for different energy resolution of detector.

It's noticeable that the energy resolution of the detector is extremely important for detecting the effect of the oscillations. It should be noted that integration of the matrix by energy or distance significantly inhibits the ability to detect the effect of oscillations. Besides, it should be noted that the measurements at 6 – 9 m is the most important, measurements at 9 - 12 m can't bring significantly contribution. But it is important for correct normalization.

**12. Measurements 2018 – 2019**

The new measurement cycle began in September 2018 and continued until July 2019. In July 2019, the reactor was stopped for reconstruction. However, background measurements were continued until December 2019. Additional background measurements on the non-operating reactor are very useful because in measurements with the operating reactor the part of background measurements was approximately 40% of the total measurement time. Since background contributes significantly to statistical error, additional background measurements have allowed for further improvement in the statistical accuracy of the experiment.

Measurements from September 2018 to July 2019 were carried out mainly in near positions to the reactor, where the ratio of the effect to background is significantly better. This made it possible to almost double the statistics collected in half the time and thus increase the statistical accuracy of measurements by 1.4 times.

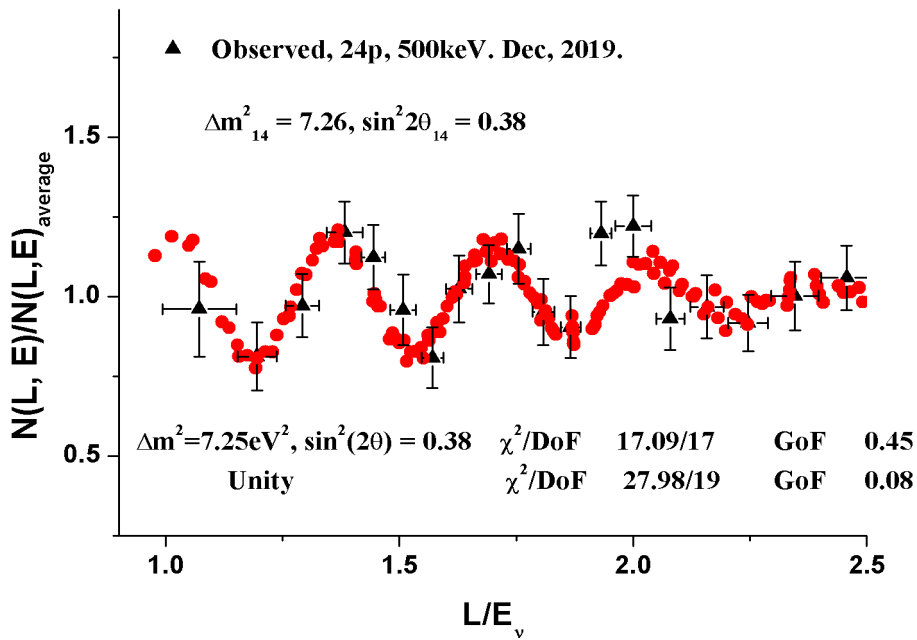

Fig. 22 The most important part of effect of antineutrino oscillation in sterile neutrino using coherent summation method.

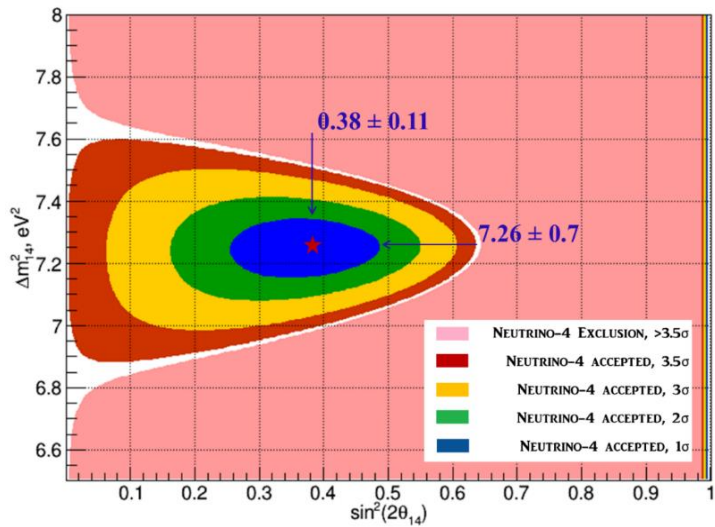

Fig. 23. a – Restrictions on parameters of oscillation into sterile state with more than 3.5σ CL (pink), area of acceptable with 3.5σ CL values of the parameters (red), area of acceptable with 3σ CL values of the parameters (yellow), area of acceptable with 2σ CL values of the parameters (green), area of acceptable with 1σ CL values of the parameters (blue).

a

The stability of the results of measurements is characterized by distributions of ON-OFF difference fluctuations normalized on their statistical uncertainties, in measurements within one period. As before the distribution has the form of a normal distribution, but its width exceeds unit by (7 ± 4.5) %. This is a result of additional dispersion which appears probably due to fluctuations of the cosmic background and impossibility of simultaneous measurements of the effect and background. Since the measurements of the background carried out during the annual scheduled reactor repair works, when the reactor is stopped for a month, are added to total obtained data, then total additional dispersion, which is a result of background measurements, increases up to 9%.

It should be noted that the validity of the spreading effect is approximately 1.6σ. The increasing of statistical measurement errors by 9% was done in the processing of experimental data. It may be an overestimation, but this guarantees the reliability of the final result.

The entire set of experimental data was processed in the same manner as previously described. The proposed method of coherent addition of results with the same value of parameter L/E gives the most visible picture of the process of oscillations The effect of the oscillations detected in the first cycle at confidence level 3 $\sigma$ is now confirmed at confidence level 3.5 $\sigma$. The effect of antineutrino oscillation in sterile neutrino is presented in Fig. 22 and Fig. 23.

The most probable parameters of oscillation are $\Delta m_{14}^2 \approx 7.26 eV^2, \sin^2 2\theta_{14} \approx 0.38 \pm 0.11$ (3.5σ). The fit with such parameters has the goodness of fit 45%, while fit with a constant equal to one (assumption of no oscillations) has the goodness of fit only 8%. We obtained $\chi^2/DOF = 17/17$ for the version with oscillation and $\chi^2/DOF = 28/19$ for the version without oscillation. In Fig. 22 the vertical errors are statistical, and the horizontal errors correspond to the interval of averaging of data.

Results of measurements of the difference in counting rate of neutrino events are shown in Fig. 24, as dependence of antineutrino flux on the distance to the reactor core. Fit of an experimental dependence with the law A/L² yields satisfactory result. Goodness of that fit is 70%. Corrections for finite size of reactor core and detector sections are negligible – 0.3%, and correction for difference between detector movement axes and direction to center of reactor core is also negligible – about 0.6%. Corrections for fast neutrons from reactor is approximately 3%.

Fig. 25 shows the famous [2] oscillation curve of the reactor antineutrino with insertion of the picture of the oscillations obtained in the Neutrino-4 experiment. The effect of gallium anomaly deficiency is shown on the left [21,22]. Here is also presented the result of the ILL experiment at a distance of 9m. Combination of these results gives an estimation for mixing angle $\sin^2 2\theta_{14} \approx 0.36 \pm 0.07$ (5.1σ), whereas the result for reactor anomaly is yielded $\sin^2 2\theta_{14} \approx 0.13 \pm 0.05$ (2.6σ).

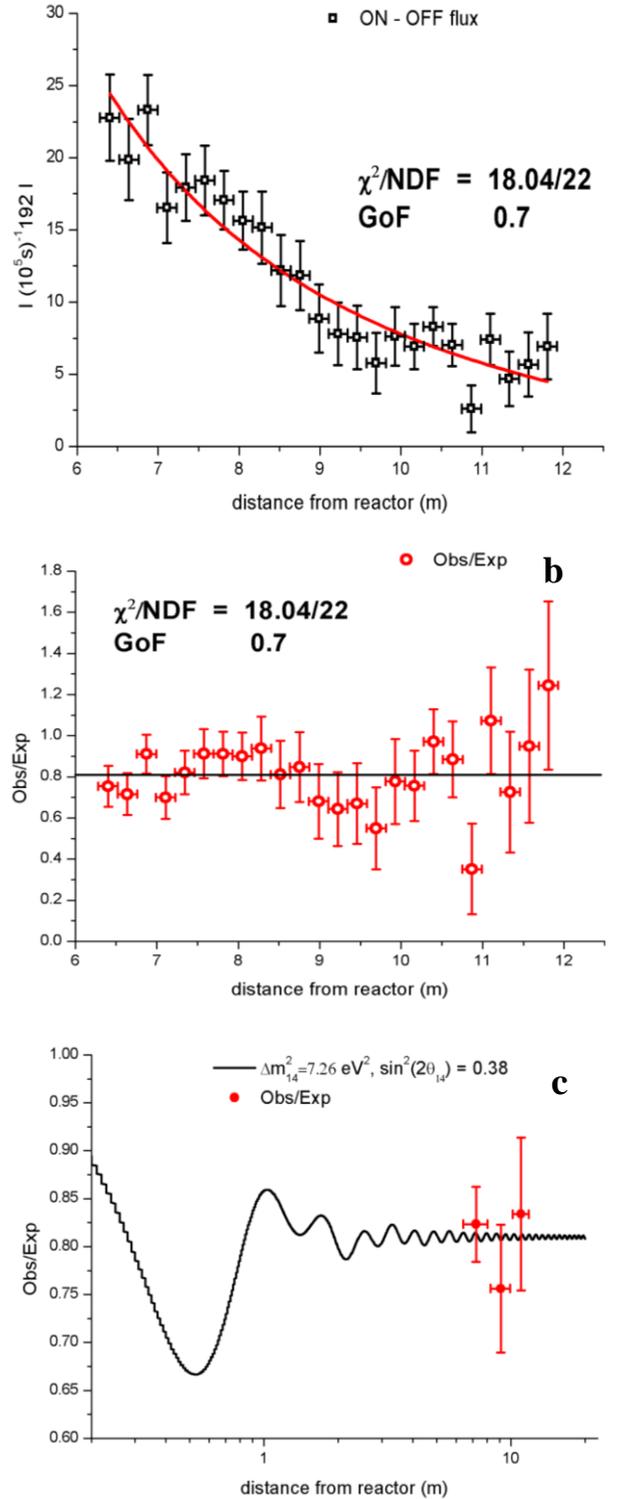

Fig. 24. Dependence of antineutrino flux on the distance to the reactor core. a - direct experimental dependence, b – normalized experimental dependence, c) oscillation curve with the experimental results in range 6-12 m.

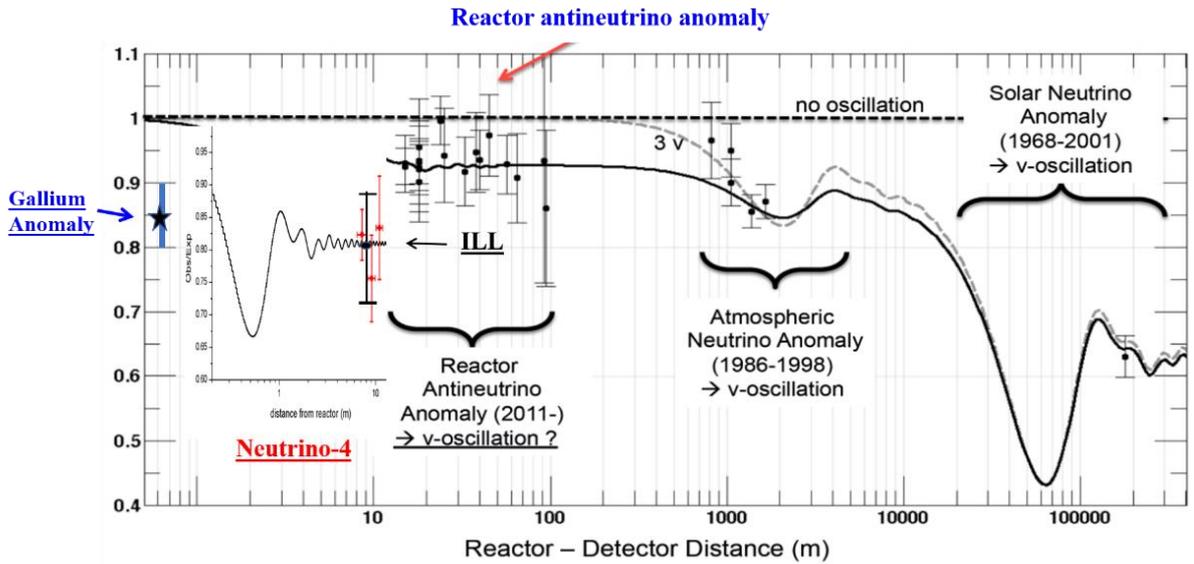

Fig. 25. Reactor antineutrino anomaly with oscillation curve from experiment Neutrino-4.

## 13. Comparison with other results of experiments at research reactors and nuclear power plants

The obtained results should be compared with other results of experiments at research reactors and nuclear power plants. Fig.26 illustrates sensitivity of other experiments NEOS [12], DANSS [16], STEREO [17] and PROSPECT [18] together with Neutrino-4.

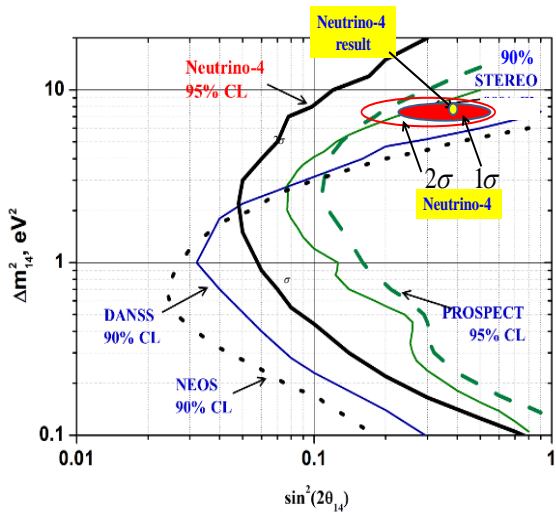

Fig. 26. Comparison of Neutrino-4 results with other experiments: – sensitivity regions of various experiments

For experiments on nuclear power plants sensitivity to identification of effect of oscillations with big $\Delta m_{14}^2$ is considerably suppressed because of the big sizes of an active zone. Experiment Neutrino-4 has some advantages in sensitivity to big values of $\Delta m_{14}^2$ owing to a compact reactor core, close minimal detector distance from the reactor and wide range of detector movements. Next highest sensitivity to large values of $\Delta m_{14}^2$ belongs to PROSPECT experiment. Currently its sensitivity is two times lower than Neutrino-4 sensitivity, but it recently has started data collection, so it possibly can confirm or refute our result. The BEST experiment started in August 2019 in BNO has good sensitivity at $\Delta m_{14}^2 > 5eV^2$ area [19]. There is combined analysis of Neutrino-4 result and BEST potential in case of gallium anomaly confirmation in this experiment [20].

It should be noted that without method of the coherent summation of data by L/E parameter, it is practically impossible to extract the effect of the oscillations. So far, the method of coherent summation of data by the parameter L/E at the short distance has been actively used only by experiment Neutrino-4 as shown in Fig. 27. This may determine the difference between the results of these experiments.

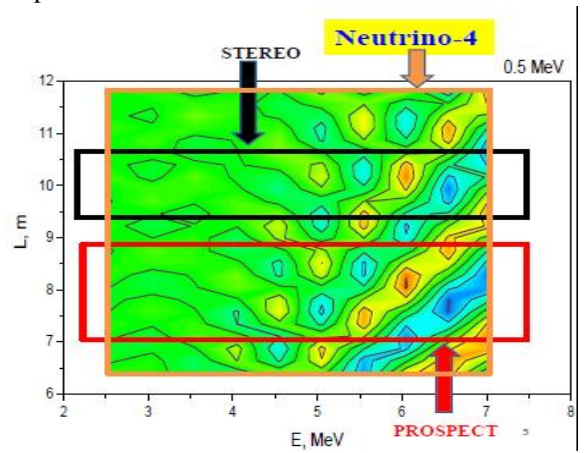

Fig. 27. Comparison of (E, L) planes for Neutrino-4, STEREO and PROSPECT.

**14. Future – new experimental installation**

Below we discuss the future prospects of Neutrino-4 experiment. Increasing of experimental accuracy is required. For that reason, the improvement of current setup and creation of new neutrino lab with new detector system at SM-3 reactor is planned.

Firstly, the improvement of current setup requires replacing of currently used scintillator with a new highly efficient liquid scintillator with capability of pulse-shape discrimination, and with an increased concentration of gadolinium up to 0.5%. It is expected that the accidental coincidence background will be reduced by factor of 3 and measurement accuracy will be doubled. Moreover, anti-coincidence shielding will be increased. The new detector is planned to be implemented with participation of colleagues from JINR and NEOS collaboration.

According to preliminary estimations, in two years of collecting data, we expect to obtain statistical accuracy at the level of 1-2% by measuring an antineutrino flux from the reactor. Thus, the question of the possible existence of a sterile neutrino with parameters of $\Delta m_{14}^2 \approx (0.5 \div 10)\text{eV}^2$ and $\sin^2(2\theta_{14}) > 0.05$ will be resolved. The accuracy of present result can reach $5\sigma$.

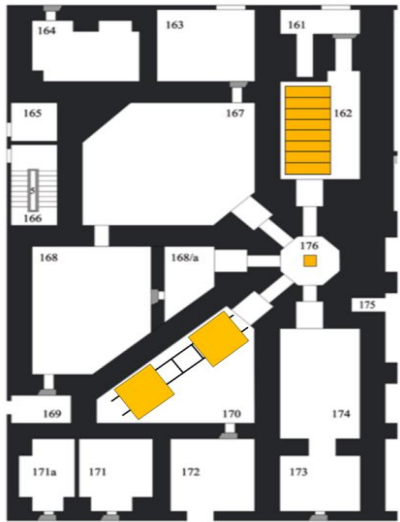
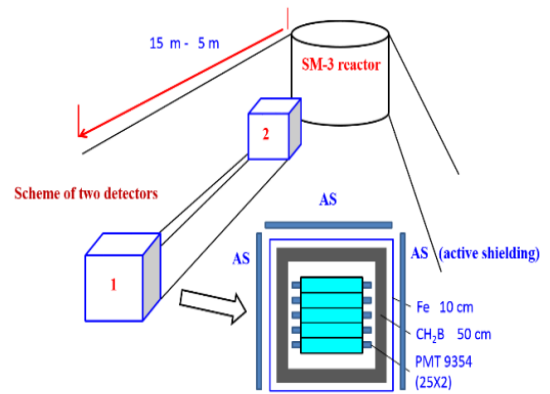

Fig. 28. The new neutrino lab with new detector system at SM-3 reactor is planned.

**15. Conclusions**

The result of presented analysis can be summarized in several conclusions. Area of reactor and gallium anomaly for $\Delta m_{14}^2 < 3\text{eV}^2$ and $\sin^2 2\theta_{14} > 0.1$ is excluded at CL more than 99.7% (>3σ).

However, we observed an oscillation effect at CL $3.5\sigma$ in vicinity of $\Delta m_{14}^2 \approx 7.26\text{eV}^2$ and $\sin^2 2\theta_{14} \approx 0.38 \pm 011$. Considering the instability of cosmic background, we increased the uncertainties of the experimental results by 9% relatively to statistical uncertainties, therefore the instability of cosmic background is already considered.

In conclusion, the current situation with the existence of sterile neutrinos should be discussed. Table 1 presents results of different experiments.

Table 1

|  | Reactor anomaly | Neutrino-4 | Gallium anomaly |
|---|---|---|---|
| $\sin^2 2\theta_{14}$ | $0.13 \pm 0.05$ ($2.6\sigma$) | $0.38 \pm 0.11$ ($3.5\sigma$) | $0.32 \pm 0.10$ ($3.2\sigma$) |
|  |  | $0.35 \pm 0.07$ ($5.0\sigma$) | |
|  | $0.2 \pm 0.04$ ($5.0\sigma$) | | |

There are three types of experiments in which a deficiency in antineutrino (neutrino) registration is observed:
1) in a number of reactor experiments, the so-called reactor anomaly,
2) in the Neutrino-4 experiment,
3) in experiments with neutrino source based on Cr51 (gallium anomaly [21,22]).

This deficiency can be interpreted as an effect of oscillations in a sterile neutrino with the mixing angle $\sin^2 2\theta_{14}$ which value is numerically the double of the observed deficiency. The values of $\sin^2 2\theta_{14}$ obtained in different experiments are presented in the Table 1. The combined result of the Neutrino-4 and gallium anomaly gives $\sin^2 2\theta_{14} \approx 0.35 \pm 0.07$ ($5.0\sigma$).

A combination of results of all experiments gives value $\sin^2 2\theta_{14} \approx 0.2 \pm 0.04$ $(5.0\sigma)$ The validity of combining the results of the reactor anomaly and the results of the Neutrino-4 experiment is questionable, since the difference in the values of these experiments is $0.25 \pm 0.12$ (2σ). However, it should be noted that the results of the reactor anomaly do not yet include a systematic calculation error, which is currently under discussion.

The difference of $\sin^2 2\theta_{14}$ for reactor and gallium anomalies makes $0.19 \pm 0.11$ (1.7σ), and at last, the difference of $\sin^2 2\theta_{14}$ for an experiment of the Neutrino-4 and gallium anomaly makes $0.06 \pm 0.15$ (0.4σ).

Nevertheless, it is possible to accept result of association of all experiments which gives mixing angle $\sin^2 2\theta_{14} \approx 0.2 \pm 0.04$ $(5.0\sigma)$. Of course, at a confidence level $5\sigma$ the existence of a sterile neutrino can be questioned. The value $\Delta m^2_{14} = (7.3 \pm 0.7) \text{eV}^2$ is obtained from the Neutrino-4 experiment with an accuracy of ~ 10%, which is determined by the accuracy of the energy calibration.

**Neutrino mass**

It is necessary to notice that obtained values $\sin^2 2\theta_{14} = (0.25 \pm 0.05)$ and $\Delta m^2_{14} = (7.3 \pm 0.7)\text{eV}^2$ allow make assessment on the mass of a neutrino: $m_\beta \approx \sqrt{7.26 \cdot 0.2}/2 \approx 0.6$.

For $\sin^2 2\theta_{14} \approx 0.35 \pm 0.07$ $(5.0\sigma)$ from experiment Neutrino-4 and from gallium anomaly $m_\beta \approx \sqrt{7.26 \cdot 0.35}/2 \approx 0.8$ eV. Obtained result on neutrino mass does not contradict the restriction on neutrino mass $m_\beta \leq 1$ eV from the KATRIN experiment [23]. Furthermore, the results of the determination of the sterile neutrino parameters make it possible to predict the value of $m_\beta \approx 0.6 \div 0.8$ eV that can be obtained in the KATRIN experiment. The calculation diagram for Neutrino-4 only is shown below.

# Experiment KATRIN and sterile neutrino from Neutrino-4

$$\Delta m^2_{14} = 7.26 eV^2, \quad Sin^2 2\vartheta_{14} = 0.38$$

$$m_\beta = \sqrt{\sum_i m_i^2 /U_{ei}/^2}$$

$$\Delta m^2_{14} \approx m_4^2,$$

$$Sin^2 2\vartheta_{14} = 4/U_{14}/^2 (1-/U_{14}/^2)$$
$$/U_{14}/^2 \ll 1$$
$$/U_{14}/^2 \approx \frac{1}{4} Sin^2 2\vartheta_{14}$$

$$m_\beta \approx \frac{1}{2}\sqrt{7.3 \cdot 0.38} \approx 0.87 eV$$

**1. There is no contradiction with restriction from experiment KATRIN -** $m_\beta \leq 1eV$

**2. If effect of Neutrino-4 is correct then prediction for neutrino mass is** $m_\beta \approx 0.87 eV$

More detailed analysis of comparison between Neutrino-4 and KATRIN gives the following estimations. Upper limit for KATRIN result 1.1eV(90%) means neutrino mass estimation $m_\beta = 0.30 \pm 0.49$eV; Neutrino-4 estimation for neutrino mass is $m_\beta = 0.87^{+0.07}_{-0.17}$eV. That means discrepancy is $\Delta m_\beta = 0.57 \pm 0.50$ (1σ) and it is not serious contradiction like it was declared in [24].

**Acknowledgements**

The authors are grateful to the Russian Foundation of Basic Research for support under Contract No. 14-22-03055-ofi_m. Authors are grateful to Y.G.Kudenko, V.B.Brudanin, V.G.Egorov, Y.Kamyshkov and V.A.Shegelsky for beneficial discussion of experimental results. The delivery of the scintillator from the laboratory headed by Prof. Jun Cao (Institute of High Energy Physics, Beijing, China) has made a considerable contribution to this research.

**References**

[1]  T. Mueller, D. Lhuillier, M. Fallot et al., Phys. Rev. C 83, 054615 (2011).

[2]  G. Mention, M. Fehner, Th. Lasserre et al., Phys. Rev. D 83, 073006 (2011).

[3]  S. Gariazzo, C. Giunti, M. Laveder and Y.F. Lie, J. High Energ. Phys. (2017) 2017: 135, arXiv:1703.00860

[4]  A. P. Serebrov, V. G. Ivochkin, R. M. Samoilov et al., Tech. Phys. 60, 1863 (2015); arXiv:1501.04740

[5]  A. P. Serebrov, V. G. Ivochkin, R. M. Samoilov et al., JETP 121, 578 (2015); arXiv:1501.04740

[6]  A. P. Serebrov, V. G. Ivochkin, R. M. Samoilov et al., Tech. Phys. 62, 322 (2017); arXiv:1605.05909

[7]  A. P. Serebrov, V. G. Ivochkin, R. M. Samoilov, et. al., arXiv:1708.00421

[8]  S.-H. Seo (RENO), Proceedings, 26th International Conference on Neutrino Physics and Astrophysics (Neutrino 2014): Boston, Massachusetts, United States, June 2-7, 2014, AIP Conf. Proc. 1666, 080002 (2015), arXiv:1410.7987 [hep-ex].

[9]  J. H. Choi, W. Q. Choi, Y. Choi, et al., Phys. Rev. Lett. 116, 211801 (2016), arXiv:1511.05849 [hep-ex].

[10] F. P. An, A. B. Balantekin, H.R. Band, et al., Phys. Rev. Lett. 116, 061801 (2016), arXiv:1508.04233 [hep-ex].

[11] Y. Abe, S. Appel, T. Abrahao, et al., JHEP 01, 163 (2016), arXiv:1510.08937 [hep-ex].

[12] Y. J. Ko, B. R. Kim, J. Y. Kim, et al., Phys. Rev. Lett 118, 121802 (2017)

[13] P. Huber, Phys. Rev. Lett. 118, 042502 (2017), arXiv:1609.03910

[14] C. Giunti, Phys. Lett. B 764, 145 (2017), arXiv:1608.04096

[15] G. Bak, J. H. Choi, H. I. Jang, et. al. (RENO Collaboration), Phys. Rev. Lett. 122, 232501 (2019) arXiv:1806.00574

[16] I. Alekseev, V. Belov, V. Brudanin, et. al. (DANSS Collaboration), Phys. Lett., B 787, 56 (2018). arXiv:1804.04046

[17] H. Almazán, L. Bernard, A. Blanchet, et. al. (STEREO Collaboration), arXiv:1912.06582

[18] J. Ashenfelter, A.B. Balantekin, C. Baldenegro, et. al. (PROSPECT Collaboration), Phys. Rev. Lett. 121, 251802 (2018). arXiv:1806.02784

[19] V. Barinov, B. Cleveland, V. Gavrin, D. Gorbunov, and T. Ibragimova, Phys. Rev. D, 97, 073001 (2018). arXiv:1710.06326

[20] V. Barinov, V. Gavrin, V. Gorbachev, D. Gorbunov, and T. Ibragimova, Phys. Rev. D 99, 111702(R) (2019)

[21] W. Hampel, J. Handt, G. Heusser, et al. (GALLEX Collaboration), Phys. Lett., B 447, 127 (1999)

[22] J. N. Abdurashitov, T.J. Bowles, M.L. Cherry, et al. (SAGE Collaboration), Phys. Rev. C, 60, 055801 (1999).

[23] M. Aker, K. Altenmuller, M. Arenz, et al. (KATRIN Collaboration), Phys. Rev. Lett. 123, 221802 (2019). arXiv:1909.06048

[24] C. Giunti, Y.F. Li, and Y.Y. Zhang. arXiv:1912.12956v2 [hep-ph]